\documentclass[aps,prl,twocolumn,groupedaddress,floatfix]{revtex4}

\usepackage{graphicx}
\usepackage{bm}
\usepackage{subfigure}
\usepackage{latexsym}
\usepackage{placeins}
\usepackage{amstext,amssymb,amsfonts}
\usepackage{epsfig}
\usepackage{color}
\usepackage{titlesec}
\usepackage{amsmath}
\usepackage{float}
\usepackage{textcase}

\begin{document}


\title{Optimal resource allocation for network functionality}
\author{Amikam Patron$^1$, Reuven Cohen$^2$}

\affiliation{1 Department of Physics, Bar-Ilan University, Ramat-Gan 5290002, Israel\\
  2 Department of Mathematics, Bar-Ilan University, Ramat-Gan 5290002, Israel}

\date{\today}

\begin{abstract}
The traditional approach to network robustness, is based on comparing network parameters before and after an event of nodes removal, such as the change in network diameter, the change in giant component size and the existence of giant component. Recently, a new approach to network robustness was presented, where the network functionality during its entire life span (during the node removal event) is considered. This approach considers nodes removal due to \textit{aging} -- a state where nodes can be degenerated and become nonfunctional, when their survival time duration -- their \textit{lifetime} -- is passed. Accordingly, a problem that has to be solved is: in the network design stage, how to divide a budget of lifetime between the network's nodes, such that the network functionality during all the stages of nodes removal due to aging, is maximized. To date, the problem has been solved only partially and numerically. In this paper we solve the problem analytically, and derive a criterion for choosing the right set of nodes on which the total lifetime budget should be divided. We also find analytically the best way of dividing the lifetime budget between the chosen set nodes, such that the network robustness with consideration to its functionality in the entire life span, is maximized.
\end{abstract}

\maketitle

\section{Introduction}
The robustness of a network, which is the capability of a network to keep being functional even when some of its nodes (or links) are removed, is an issue that has been researched and studied widely \cite{cohen-prl-2000,callaway-prl-2000,cohen-prl-2001,pastor-pre-2001,newman-arxiv-2002,paul-europ-phys-journal-2004,gallos-prl-2005,cohen-2010,morone-nature-2015,braunstein-pnas-2016,buldyrev-nature-2010,buldyrev-pre-2011,albert-nature-2000,kim-prl-2003,motter-pre-2002,holme-pre-2002,albert-pre-2004,holme-pre-2007,shargel-prl-2003,jahnke-prl-2008,Schneider-proc-nat-acad-sci-2011,hermann-jrnl-stast-mech-2011,lin-epl-2018}. Traditionally, the measurement of network robustness has been performed in states where a removal of nodes is a result of an \textit{attack} on the network. There are two types of attacks, in which the research has been focused. The first type is \textit{random attack}, where the choice of the network's nodes for removal is performed randomly, without preference of a node to be removed over any other node in the network. The second type is \textit{targeted attack} where, in contrast to random attack, the attack is targeted against the nodes with the central role of maintaining the connectivity of the network.

There have been many studies about network robustness, that are based on percolation theory \cite{cohen-prl-2000,callaway-prl-2000,cohen-prl-2001,pastor-pre-2001,newman-arxiv-2002,paul-europ-phys-journal-2004,gallos-prl-2005,cohen-2010,morone-nature-2015,braunstein-pnas-2016}. According to this approach, there are two states in which a network can be found -- the \textit{subcritical state} where the network is fragmented to many small components each of them contains relatively small number of nodes, and the \textit{supercritical state} where there exists a relative big component which contains a finite fraction of the entire network's nodes, i.e. scales as $O(N)$, named the \textit{giant component}, and aside from it possibly exist small components. The transition between the supercritical state to the subcritical state, where the giant component is fragmented into small components, occurs when a critical combination of the network's nodes is removed. The fraction of the network's nodes that has not to be removed in order to guarantee the existence of the giant component, is named the \textit{percolation threshold} of the network, denoted by $p_c$.

In \cite{cohen-prl-2000,callaway-prl-2000,cohen-prl-2001}, the percolation thresholds for two kinds of random networks that have been studied widely, were measured. For Erd\H{o}s-R\'{e}nyi (ER) network, which was the first model of random network to be studied \cite{erdds1959random,erd6s1960evolution}, where the node's degree follows a Poisson distribution $P(k)=e^{-\lambda}\frac{\lambda^k}{k!}$, and most of the nodes' degrees are close to the expected degree $\lambda$, it was found that under random attack the percolation threshold is $p_c=\frac{1}{\lambda}$. For Scale-Free (SF) network, a topology that was found in many real networks \cite{Redner-eurp-phs-j-1998,albert-nature-1999,Barabasi-science-1999,Newman-pnas-2001}, where the node's degree follows power-law distribution $P(k)\sim k^{-\gamma}$, such that beside most of the nodes that have a very small degree there are some nodes with a very high degree that are named the \textit{hubs} of the network, it was found that under random attack, for $\gamma>3$, $p_c$ has a finite value, but for $\gamma\leq3$, $p_c$ approaches $0$. This means that even when almost all of the nodes are removed, the network is still functional and a giant component exists in the network. For targeted attack it was found that for ER network, $p_c$ is close to its value under random attack, since most of the nodes' degrees are close to $\lambda$. In contrast, for SF network it was found that under targeted attack $p_c$ is high, since removing only the very small group of the network's hubs, that have a critical role of maintaining the network connectivity, causes the network to be fragmented. The percolation model for network robustness, was also generalized for interdependent networks \cite{buldyrev-nature-2010,buldyrev-pre-2011}. 

In addition to measuring network robustness by the percolation threshold, there have been other parameters which were defined for measuring network robustness, like change in the \textit{network diameter} which is the average short path length between two randomly chosen nodes \cite{albert-nature-2000,shargel-prl-2003,kim-prl-2003}, the relative size of the giant component which is the ratio between the giant component size after and before the attack \cite{motter-pre-2002}, change in the \textit{betweenness centrality} which is the average of number of shortest paths between pairs of the network's nodes passing through randomly chosen node \cite{holme-pre-2002}, \textit{connectivity loss} which is the average of the ratio between the number of nodes connected to randomly chosen node after and before the nodes removal \cite{albert-pre-2004}, and \textit{f-robustness} which is the number of nodes to be removed such that the giant component size is reduced by factor $f$ ($f\in[0,1]$) \cite{holme-pre-2007}.

Efforts also have been made to find a topology for random network on which the robustness is optimized both for random and targeted attack, based on diameter change \cite{shargel-prl-2003} or percolation threshold \cite{paul-europ-phys-journal-2004}. In addition to the studies about random networks robustness, there are also some studies about robustness of real networks like the North American power grid \cite{albert-pre-2004} and optical communication networks \cite{jahnke-prl-2008}. 

A new approach to network robustness has been proposed in \cite{Schneider-proc-nat-acad-sci-2011,hermann-jrnl-stast-mech-2011}. According to it, a measurement of robustness should consider the whole life span of a network. This way is different from the traditional methods, that measure robustness by comparing a final and initial states after and before an attack on the network took place, respectively, with no consideration to the intermediate state of the network functionality during the attack. According to the new approach, network robustness is measured by the integral value of the fraction of nodes in the giant component with respect to the number of nodes were removed from the network, during all the stages of the network attack. In \cite{Schneider-proc-nat-acad-sci-2011,hermann-jrnl-stast-mech-2011} it was shown by simulations, that choosing a small combination of pairs of the network's links and swapping the connections of each pair (for example, the pair consisting of a link between nodes $i$ and $j$ and a link between nodes $k$ and $l$, is swapped and becomes a link between nodes $i$ and $k$ and a link between nodes $j$ and $l$), enhances the network robustness, i.e. the integral value mentioned above, very significantly. 

This approach was further elaborated in \cite{lin-epl-2018}, where the effect of \textit{node aging} has been added, as a factor that should be considered in measuring network robustness. It was argued, that each node has a typical lifetime, such that from the beginning of the network life, after the node lifetime is passed, the node becomes degenerate and in fact is removed from the network. Therefore, node aging is another cause to node removal, in addition to the traditional cause of an attack on the network. Accordingly, a network survivability function, which is the integral of the giant component size with respect to time during the entire life span of the network $T$, has been defined 
\begin{equation}
I=\int_0^T G(t)dt,
\label{eq:survivability}
\end{equation}
that is very similar in logic to the robustness integral in \cite{Schneider-proc-nat-acad-sci-2011,hermann-jrnl-stast-mech-2011} as was described above. 

For maximizing the network robustness according to it, the following problem has to be solved: in the network design stage where the designer allocates the nodes with lifetime, given a total budget of lifetime for all the network's nodes, what is the best way of dividing the lifetime budget between them, such that the survivability $I$ in Eq. (\ref{eq:survivability}) is maximized? Indeed, in \cite{lin-epl-2018} it was shown numerically, that when a node's lifetime is taken from a distribution with expectation that equals $k^\alpha$, where $k$ is the node's degree, then a critical value $\alpha_c$ can be found, dependent, among others, on the network topology, such that determining the distribution expectation to be $k^{\alpha_c}$, produces the optimized lifetime division with it the network survivability is maximized. 

However, all the previous results are based on numerical solutions or on simulations only as mentioned above, and to date there is still no analytic solution for a criterion or an algorithm according to it the network survivability would be maximized. In this paper we develop analytically a solution for the problem of network survivability maximization. We develop analytically a criterion for choosing the right set of nodes on which the total budget of lifetime should be divided, and also we find analytically the best way of dividing the lifetime between the nodes of the chosen set, in such a way that the network survivability is maximized.  

\section{Model}
\begin{figure}[t]
\begin{center}
  \begin{tabular}{c}
  \includegraphics[width=1\columnwidth]{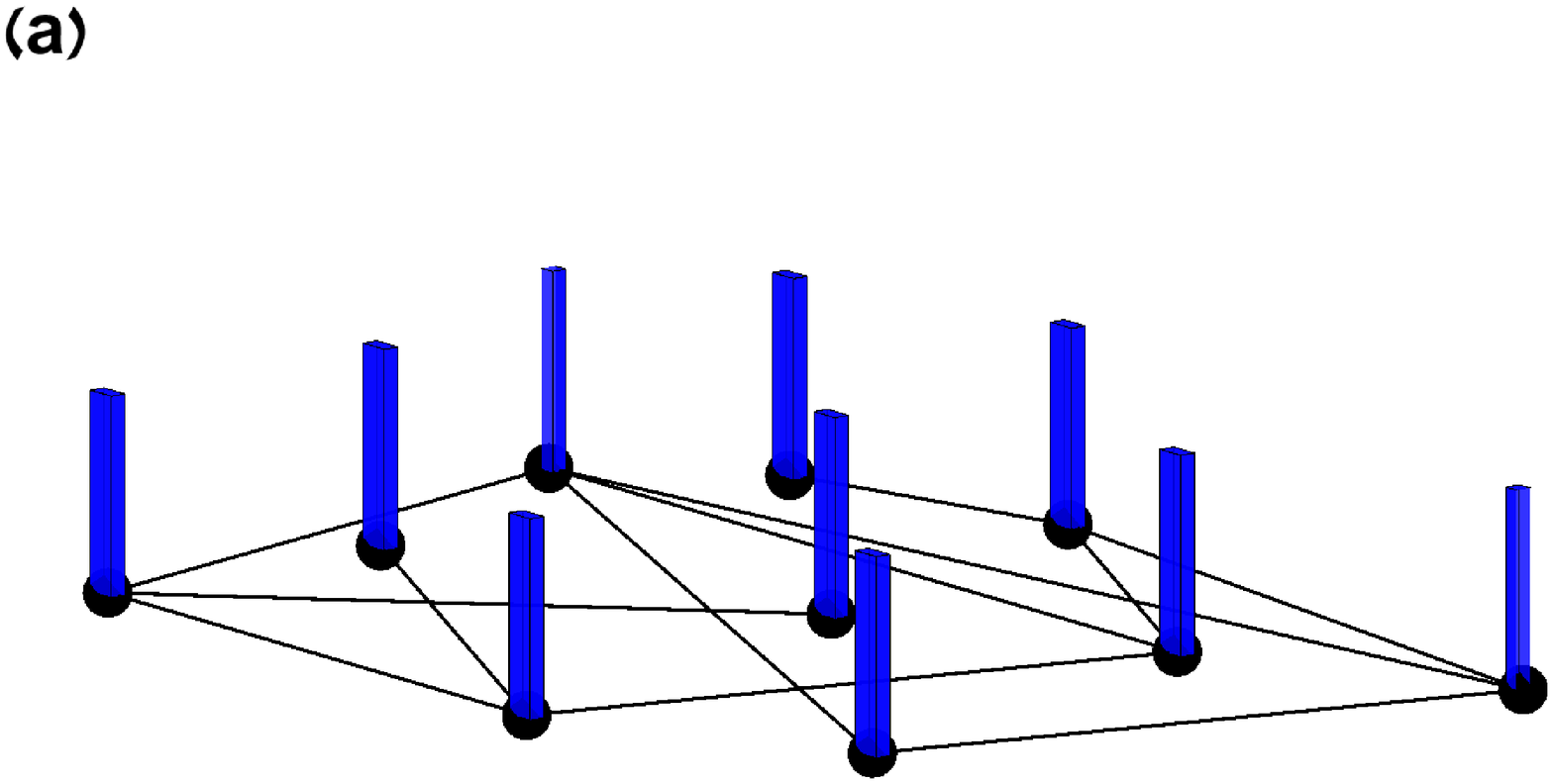}\vspace{-2.2em}\\
	\includegraphics[width=1\columnwidth]{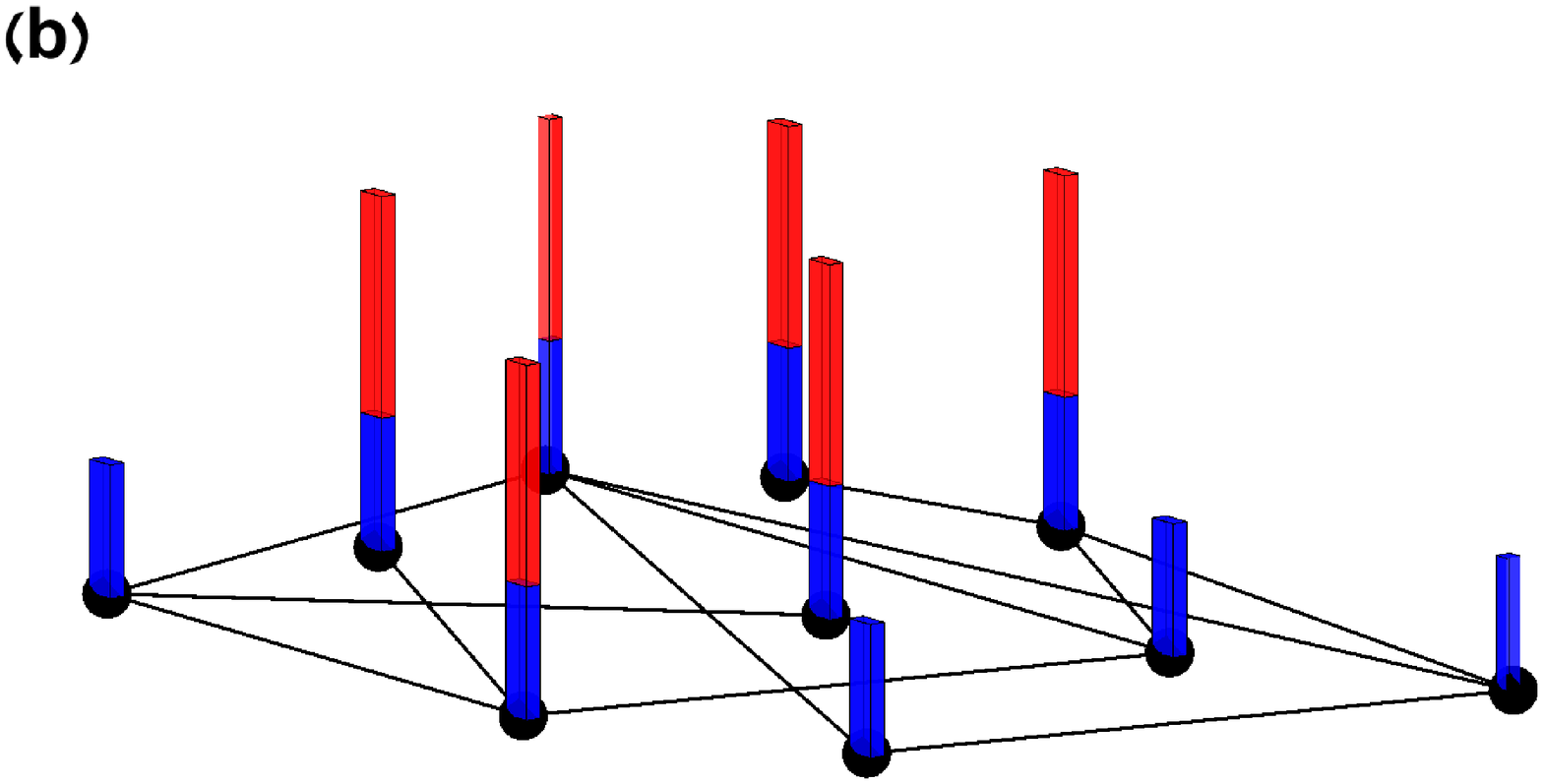}\vspace{-2.2em}\\
	\includegraphics[width=1\columnwidth]{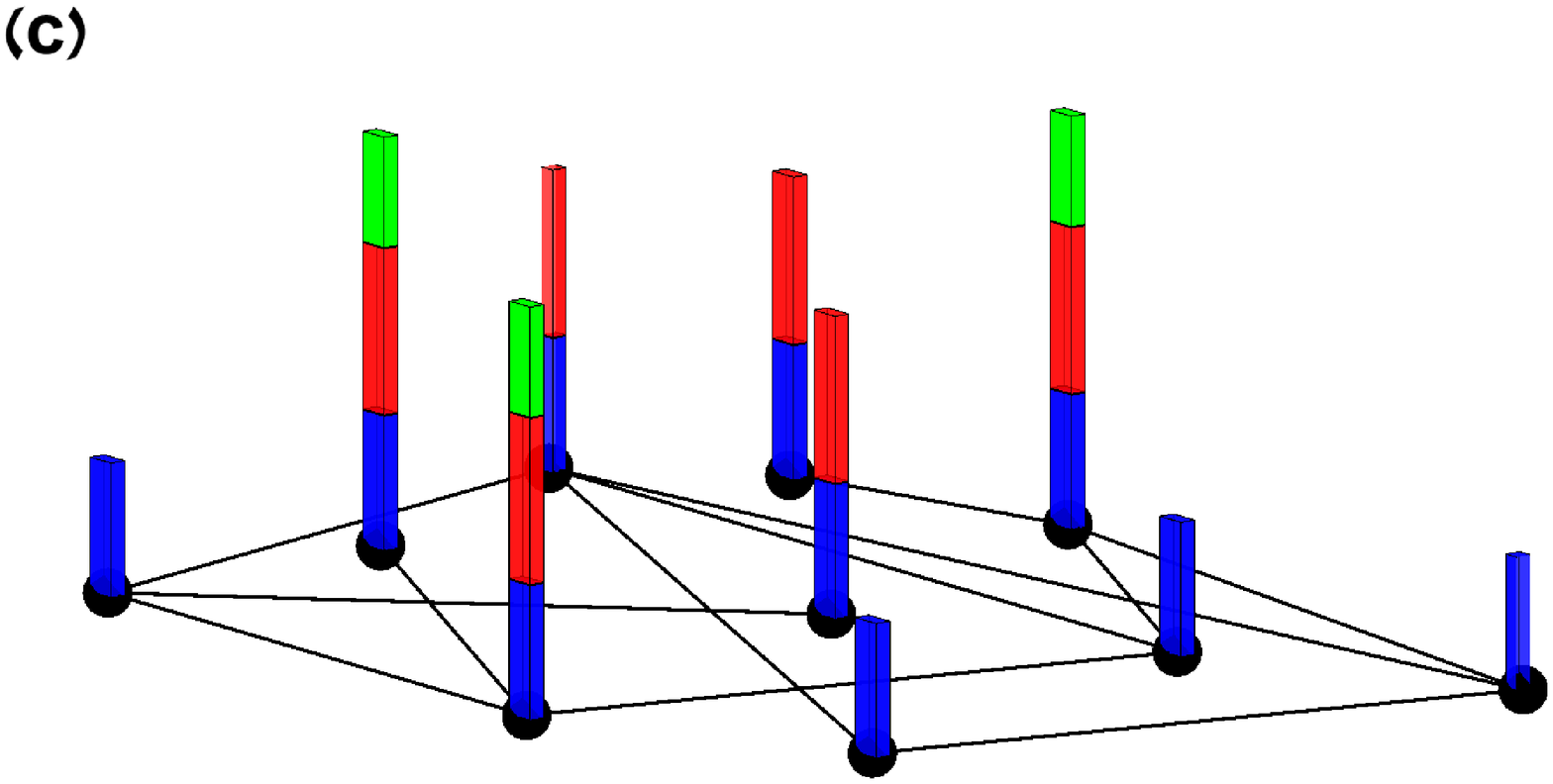}\vspace{-2.5em}\\
\end{tabular}
\end{center}
\caption{\textbf{Basic model illustration for choice of three sets $A\supset B\supset C$: (a)} Uniform division of the entire lifetime budget between the nodes of set $A$. Node's lifetime is represented by the height of the blue bars. \textbf{(b)} Transferring lifetime from set $A-B$ to set $B$. Set $B$ nodes are those with the red bars above the blue bars. Red bars height represents the lifetime of set $B$ nodes, after the lifetime transfer to this set was performed. Blue bars height represents the remained lifetime of set $A$. \textbf{(c)} Transferring lifetime from set $B-C$ to set $C$. Set $C$ nodes are those with the green bars above the red and blue bars. Green bars height represents the lifetime of set $C$ nodes after the lifetime transfer to this set was performed.}
\label{fig:basic_illustration}
\end{figure}

Our model begins with a network with $N$ nodes, and a total budget of lifetime denoted by $T_0$. We choose a finite descending sequence of sets of the network nodes $A\supset B\supset C\supset D\supset E\supset F...$ . Firstly we divide the entire lifetime budget $T_0$ uniformly between set $A$ nodes. Then we subtract an equal part of lifetime units from each of the nodes of set $A-B$ -- nodes that belong to set $A$ but not belong to set $B$, and transfer the entire lifetime was subtracted uniformly between set $B$ nodes. Then we subtract an equal part of lifetime units from each of the nodes of set $B-C$, and transfer it uniformly between set $C$ nodes. We continue this lifetime transfer in the same way from set $C-D$ to set $D$, from set $D-E$ to set $E$ etc., until we transfer lifetime to the smallest set was chosen. Fig. \ref{fig:basic_illustration} is a basic illustration of this model.

\section{Theory}
\subsection{Presentation of network survivability}
For readability, we use $\Psi=\langle A,B,C...\rangle$ to note the sets were chosen, where $\Psi_1=A, \Psi_2=B$ etc. The number of nodes in set $\Psi_i$ and set $\Psi_i-\Psi_{i+1}$, would be noted $n_{\Psi_i}$ and $n_{\Psi_i-\Psi_{i+1}}$, respectively. The lifetime of set $\Psi_i$ would be noted $t_{\Psi_i}$. The giant component size of set $\Psi_i$, when set $\Psi_i$ is the only set whose nodes are functional, would be noted $S_{\Psi_i}$. The lifetime subtracted from set $\Psi_i-\Psi_{i+1}$ and transferred to set $\Psi_{i+1}$, would be noted $\delta T_{\Psi_i-\Psi_{i+1}}$.

In general, the network survivability, with accordance to its definition in Eq. (\ref{eq:survivability}), is 
\begin{equation}
I=S_At_A+S_Bt_B+S_Ct_C+...=\sum_i S_{\Psi_i} t_{\Psi_i}.
\label{eq:base_survivability}
\end{equation}
In our model, each stage of lifetime transfer between sets, changes the parameters of Eq. (\ref{eq:base_survivability}) and thus changes the network survivability. Due to that, in order to find the final network survivability after the completion of lifetime transfers, we calculate below the survivability in stages, where in each stage we consider another lifetime transfer in the sequence of transfers.

We begin firstly with the first stage of dividing the entire lifetime budget $T_0$ uniformly between set $A$ nodes, such that $t_A=\frac{T_0}{n_A}$. Therefore the network survivability is
\begin{equation}
I=T_0\frac{S_A}{n_A}.
\end{equation}
Due to the first subtraction of $\delta T_{A-B}$ lifetime units from each of the nodes of set $A-B$, the lifetime of set $A$ nodes is reduced by $\delta T_{A-B}$ and becomes $t_A=\frac{T_0}{n_A}-\delta T_{A-B}$, and the lifetime of set $B$ nodes increases by $\delta T_{A-B}$ lifetime units over set $A$ nodes. Moreover, due to the transfer of the entire lifetime $\delta T_{A-B}n_{A-B}$ that was subtracted from set $A-B$, each node in set $B$ receives in addition $\frac{\delta T_{A-B}n_{A-B}}{n_B}$ lifetime units. Therefore, the total lifetime of set $B$ nodes is $t_B=\delta T_{A-B}\frac{n_A}{n_B}$ (the summation of $\delta T_{A-B}$ and $\delta T_{A-B}\frac{n_{A-B}}{n_B}$). Therefore, The network survivability after the first lifetime transfer was performed is
\begin{align}
I&=S_A\left(\frac{T_0}{n_A}-\delta T_{A-B}\right)+S_B\delta T_{A-B}\frac{n_A}{n_B}\nonumber\\
&=T_0\frac{S_A}{n_A}+\delta T_{A-B}n_A\left(\frac{S_A}{n_A}-\frac{S_B}{n_B}\right).
\end{align}
In similar way, due to the second transfer of $\delta T_{B-C}$ units of lifetime from each of the  nodes of set $B-C$ to set $C$, we get $t_B=\delta T_{A-B}\frac{n_A}{n_B}-\delta T_{B-C}$, and $t_C=\delta T_{B-C}\frac{n_B}{n_C}$, and the network survivability is
\begin{align}
I=T_0\frac{S_A}{n_A}+\delta T_{A-B}n_A&\left(\frac{S_B}{n_B}-\frac{S_A}{n_A}\right)+\nonumber\\
&+\delta T_{B-C}n_B\left(\frac{S_C}{n_C}-\frac{S_B}{n_B}\right).
\end{align}

In general, after the completion of a lifetime transfer $\delta T_{\Psi_i-\Psi_{i+1}}$ from each of the nodes of set $\Psi_i-\Psi_{i+1}$ to set $\Psi_{i+1}$, we get $t_{\Psi_i}=\delta T_{\Psi_{i-1}-\Psi_i}\frac{n_{\Psi_{i-1}}}{n_{\Psi_i}}-\delta T_{\Psi_i-\Psi_{i+1}}$, and $t_{\Psi_{i+1}}=\delta T_{\Psi_i-\Psi_{i+1}}\frac{n_{\Psi_i}}{n_{\Psi_{i+1}}}$. 

The final network survivability after the last lifetime transfer, to the smallest set was chosen, is completed, is
\begin{equation}
I=T_0\frac{S_{\Psi_1}}{n_{\Psi_1}}+\sum_i \delta T_{\Psi_i-\Psi{i+1}}n_{\Psi_i}\left(\frac{S_{\Psi_{i+1}}}{n_{\Psi_{i+1}}}-\frac{S_{\Psi_i}}{n_{\Psi_i}}\right).
\label{eq:survivability_general}
\end{equation}

\begin{figure}[t]
\begin{center}
  \begin{tabular}{c}
  \includegraphics[width=1\columnwidth]{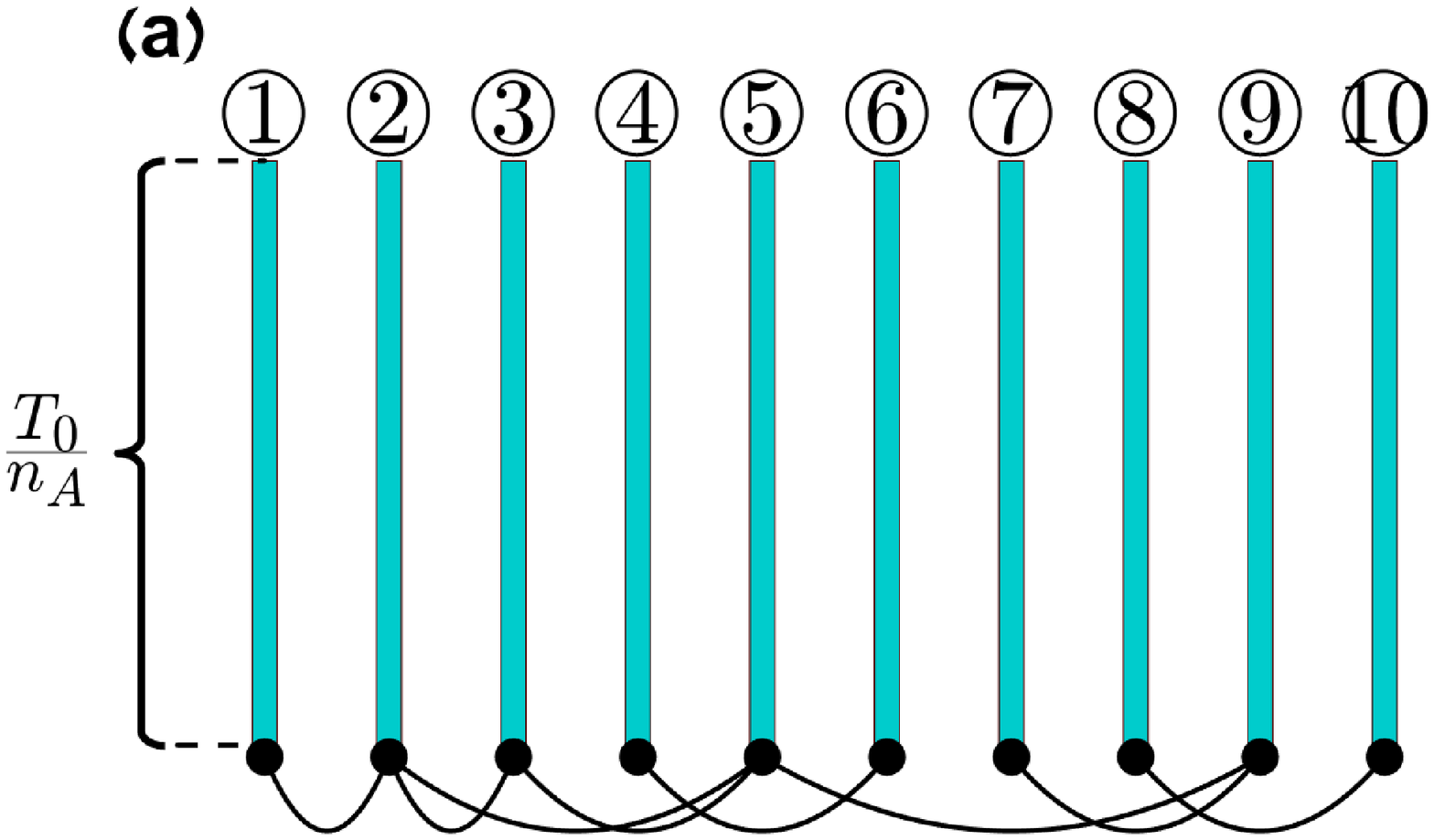}\vspace{-2.2em}\\
	\includegraphics[width=1\columnwidth]{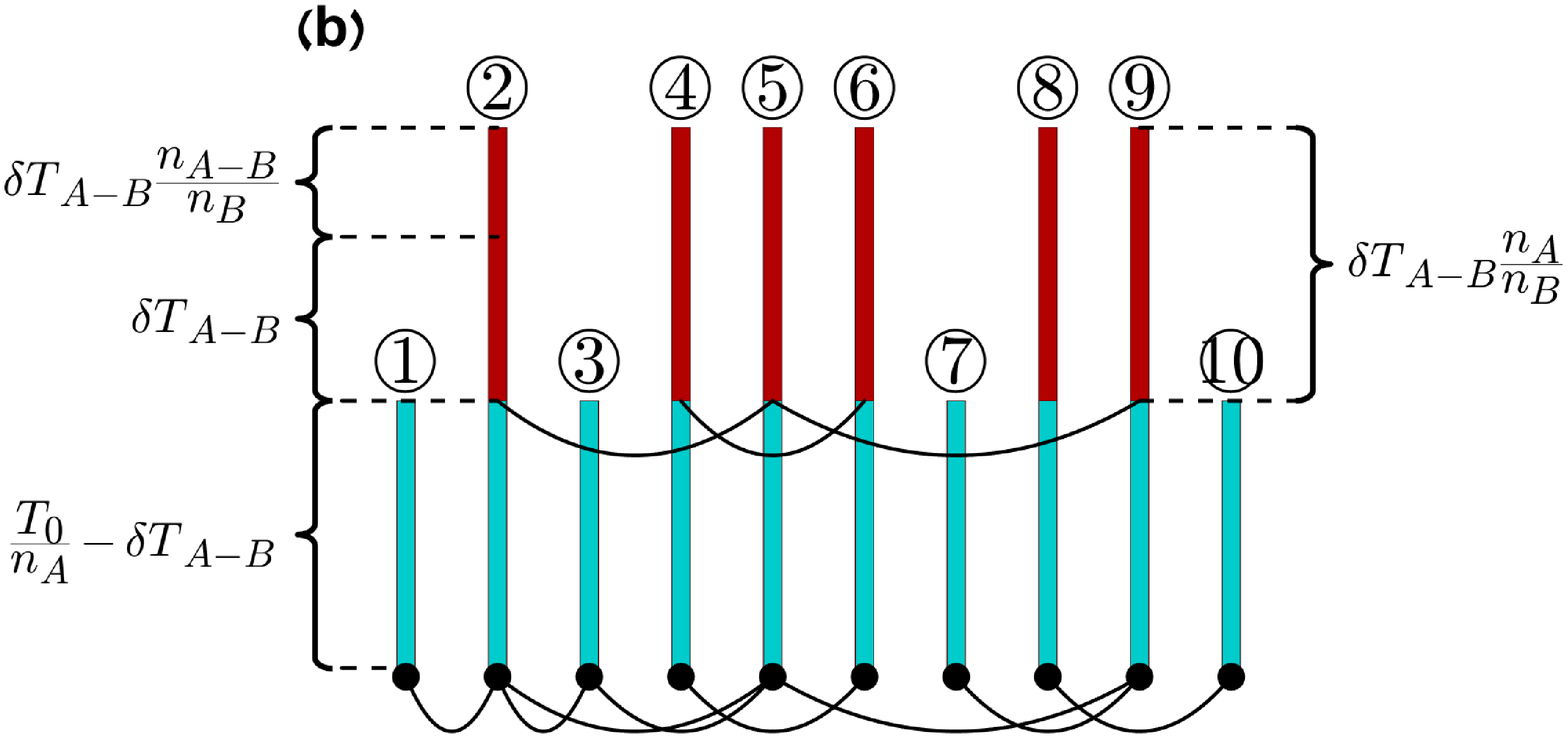}\vspace{-1.5em}\\
	\includegraphics[width=1\columnwidth]{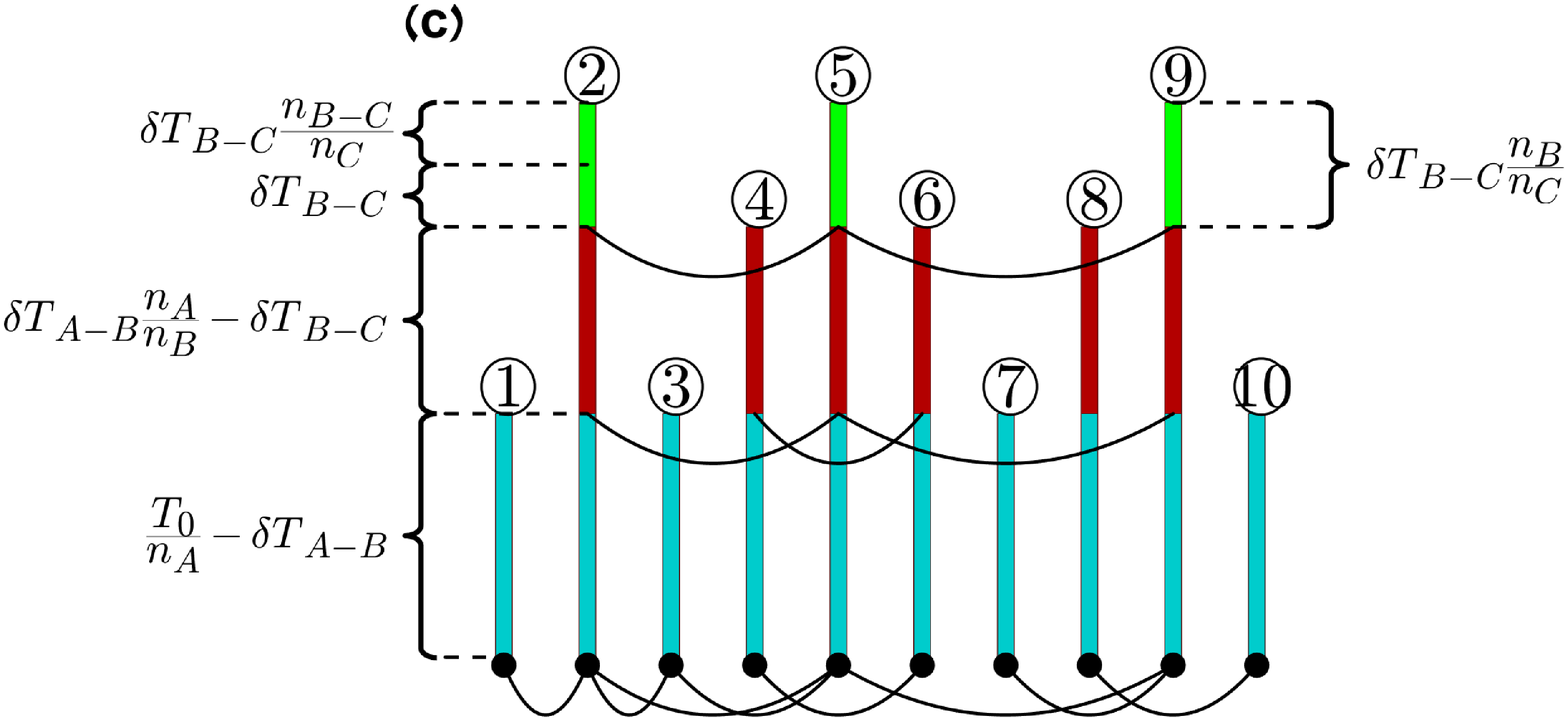}\vspace{-2.5em}\\
\end{tabular}
\end{center}
\caption{Model illustration for network with 10 nodes. Nodes are signed by numbers form '1' to '10'.\textbf{(a)} First stage -- The entire lifetime budget $T_0$ is divided uniformly between set $A$ nodes (in this case all the network's nodes). Lifetime of set $A$ -- $\frac{T_0}{n_A}$ -- is represented by the height of blue bars attached to its nodes. Links between the nodes are drawn at the bottom of the figure. \textbf{(b)} Second stage -- subtraction of $\delta T_{A-B}$ lifetime units from each of the nodes of set $A-B$ -- nodes number '1','3','7' and '10'. Set $A$ remains with $T_0-\delta T_{A-B}$ lifetime units, and set $B$ -- nodes number '2','4','5','6','8' and '9' -- gains automatically $\delta  T_{A-B}$ lifetime units greater than set $A$. Nodes in set $B$ receive also $\delta T_{A-B}\frac{n_{A-B}}{n_B}$ lifetime units, due to the lifetime transfer from set $A-B$. Total lifetime of set $B$ nodes is $t_B=\delta T_{A-B}\frac{n_A}{n_B}$. Lifetime of set $B$ is represented by the height of red bars attached to its nodes. \textbf{(c)} Third stage -- similar to (b), but when lifetime transfer is performed from set $B-C$ (nodes number '4','6' and '8') to set $C$ (nodes number '2','5' and '9'). Lifetime of set $C$ is represented by the height of green bars attached to its nodes.}
\label{fig:10nodes_model}
\end{figure}

Figure \ref{fig:10nodes_model} is an illustration of the first three lifetime transfers as described above, for a network contains $10$ nodes.

For the sake of convenience, from now on we develop the theory for only four sets $A\supset B\supset C\supset D$, but the generalization to any number of sets is simple as was shown above. Since $\delta T_{A-B}$ is subtracted from the original set $A$ lifetime $\frac{T_0}{n_A}$, and $\delta T_{B-C}$ is subtracted from the original set $B$ lifetime $\delta T_{A-B}\frac{n_A}{n_B}$, and $\delta T_{C-D}$ is subtracted from the original set $C$ lifetime $\delta T_{B-C}\frac{n_B}{n_C}$, therefore,
\begin{align}
0\leq&\delta T_{A-B}\leq\frac{T_0}{n_A}\nonumber\\
0\leq&\delta T_{B-C}\leq\delta T_{A-B}\frac{n_A}{n_B}\nonumber\\
0\leq&\delta T_{C-D}\leq\delta T_{B-C}\frac{n_B}{n_C}.
\label{eq:delta_limits}
\end{align}
We define $0\leq\alpha_A,\alpha_B,\alpha_C\leq1$. With Eq. (\ref{eq:delta_limits}) we get
\begin{align}
&\delta T_{A-B}n_A=\alpha_AT_0,\nonumber\\
&\delta T_{B-C}n_B=\alpha_B\delta T_{A-B}n_A=\alpha_B\alpha_AT_0,\\
&\delta T_{C-D}n_C=\alpha_C\delta T_{B-C}n_B=\alpha_C\alpha_B\alpha_AT_0.\nonumber
\label{eq:delat_limits2}
\end{align}
$\alpha_A$ is the fraction of set $A-B$ original lifetime $\frac{T_0}{n_A}$, was transferred to set $B$. $\alpha_B$ is the fraction of set $B-C$ original lifetime $\delta T_{A-B}\frac{n_A}{n_B}$, was transferred to set $C$. $\alpha_C$ is the fraction of set $C-D$ original lifetime $\delta T_{B-C}\frac{n_B}{n_C}$, was transferred to set $D$. 
Based on Eq. (\ref{eq:survivability_general}), the network survivability is
\begin{align}
I&=T_0\left[\frac{S_A}{n_A}+\alpha_A\left(\frac{S_B}{n_B}-\frac{S_A}{n_A}\right)\right.\nonumber\\
&\left.+\alpha_B\alpha_A\left(\frac{S_C}{n_C}-\frac{S_B}{n_B}\right)+\alpha_C\alpha_B\alpha_A\left(\frac{S_D}{n_D}-\frac{S_C}{n_C}\right)\right].
\end{align}
For each set we define $Z\equiv\frac{S}{n}$, as the fraction between its giant component size and its nodes number. We also define $U_A\equiv Z_B-Z_A$, $U_B\equiv Z_C-Z_B$ and $U_C\equiv Z_D-Z_C$. Accordingly, we get a final expression for the network survivability
\begin{equation}
I=T_0\left[Z_A+\alpha_A\left(U_A+\alpha_B\left(U_B+\alpha_CU_C\right)\right)\right].
\label{eq:survivability_U}
\end{equation}

\subsection{Maximization of network survivability}
According to Eq. (\ref{eq:survivability_U}), maximum network survivability would be achieved by maximizing the following
\begin{equation}
max\{\alpha_A\left(U_A+\alpha_B\left(U_B+\alpha_CU_C\right)\right)\}.
\label{exp:max_survivability}
\end{equation}
The maximization is implemented by determining the $\alpha$'s values from inside out, according to the rule: if the expression multiplied by $\alpha$ is greater than $0$, then $\alpha=1$, and if it is less or equal to $0$, then $\alpha=0$. 

We introduce now a specific example: firstly we have to maximize $\alpha_CU_C$. Assume that $U_C>0$ (which means $Z_D-Z_C=\frac{S_D}{n_D}-\frac{S_C}{n_C}>0$), then $\alpha_C$ is determined to be $1$, and $\alpha_CU_C$ becomes $U_C$. Accordingly, The next expression to be maximized is $\alpha_B\left(U_B+U_C\right)$. Assume that $U_B+U_C\leq0$ (which means $Z_C-Z_B+Z_D-Z_C=Z_D-Z_B=\frac{S_D}{n_D}-\frac{S_B}{n_B}\leq0$), then $\alpha_B$ is determined to be $0$, and $\alpha_B\left(U_B+U_C\right)$ becomes $0$. Accordingly, the last expression that has to be maximized is $\alpha_AU_A$. Assume that $U_A>0$ (which means $Z_B-Z_A=\frac{S_B}{n_B}-\frac{S_A}{n_A}>0$), then $\alpha_A$ is determined to be $1$, and $\alpha_AU_A$ becomes $U_A$. Therefore, according to Eqs. (\ref{eq:survivability_U}) and (\ref{exp:max_survivability}) we get for the maximum of network survivability
\begin{equation}
I_{max}=T_0\left(Z_A+U_A\right)=T_0Z_B=T_0\frac{S_B}{n_B},
\end{equation}
that means that maximum survivability is achieved by allocating the entire lifetime $T_0$, uniformly between the nodes of set $B$. Note also that among the four sets, the ratio $\frac{S}{n}$ is maximal for set $B$ (as was shown $Z_D>Z_C$ , $Z_B\geq Z_D$ and $Z_B>Z_A$), and this is the reason for choosing this set to be allocated by the entire lifetime.

The previous result can be presented in another way. According to Eqs. (\ref{eq:base_survivability}) and (\ref{eq:delat_limits2}), we get for the sets' lifetimes the followings
\begin{align}
&t_A=\left(1-\alpha_A\right)\frac{T_0}{n_A},\nonumber\\
&t_B=\alpha_A\left(1-\alpha_B\right)\frac{T_0}{n_B},\nonumber\\
&t_C=\alpha_A\alpha_B\left(1-\alpha_C\right)\frac{T_0}{n_C},\nonumber\\
&t_D=\alpha_A\alpha_B\alpha_C\frac{T_0}{n_D}.
\end{align}
Thus, we can define a vector 
\begin{equation}
\langle\hspace{0.25ex}1-\alpha_A\hspace{0.25ex},\hspace{0.25ex}\alpha_A\left(1-\alpha_B\right)\hspace{0.25ex},\hspace{0.25ex}\alpha_A\alpha_B\left(1-\alpha_C\right)\hspace{0.25ex},\hspace{0.25ex}\alpha_A\alpha_B\alpha_C\hspace{0.25ex}\rangle.\nonumber
\end{equation}
The vector components represent the lifetime of sets $A,B,C$ and $D$, respectively, as a fraction of $T_0$. With the previous rule, that maximum of the network survivability is achieved by determining the $\alpha$'s values to be be $0$ or $1$, we get that all the vector components equal $0$, except one component that equals $1$ -- the component that related to the set in which the ratio $\frac{S}{n}$ is maximum. 
For the previous example -- with $\alpha_A=1$, $\alpha_B=0$ and $\alpha_C=1$, the vector becomes $(0,1,0,0)$, where only the second component, related to set $B$, equals $1$. This presents, as was concluded above, that in order to maximize the network survivability, we have to allocate uniformly the entire lifetime to set $B$ nodes only. The above example of four sets, can be generalized to any final number of sets, performing the same analysis of network survivability as was implemented above. 

Let us summarize the previous analysis with the general rule we proved: for each given group of sets of network's nodes, all are potentially to be allocated with lifetime, the followings has to be performed in order to maximize the network survivability:
\begin{enumerate}
  \item Choice of the set in which the ratio between the giant component size and the number of nodes, is maximal.  
	\item Uniform division of the entire lifetime budget, between the nodes in the chosen set.
\end{enumerate}
      
\section{Simulations and Results}

\begin{figure*}[htbp]
\begin{center}
  \begin{tabular}{cc}
	\hspace{-7em} \includegraphics[width=10cm]{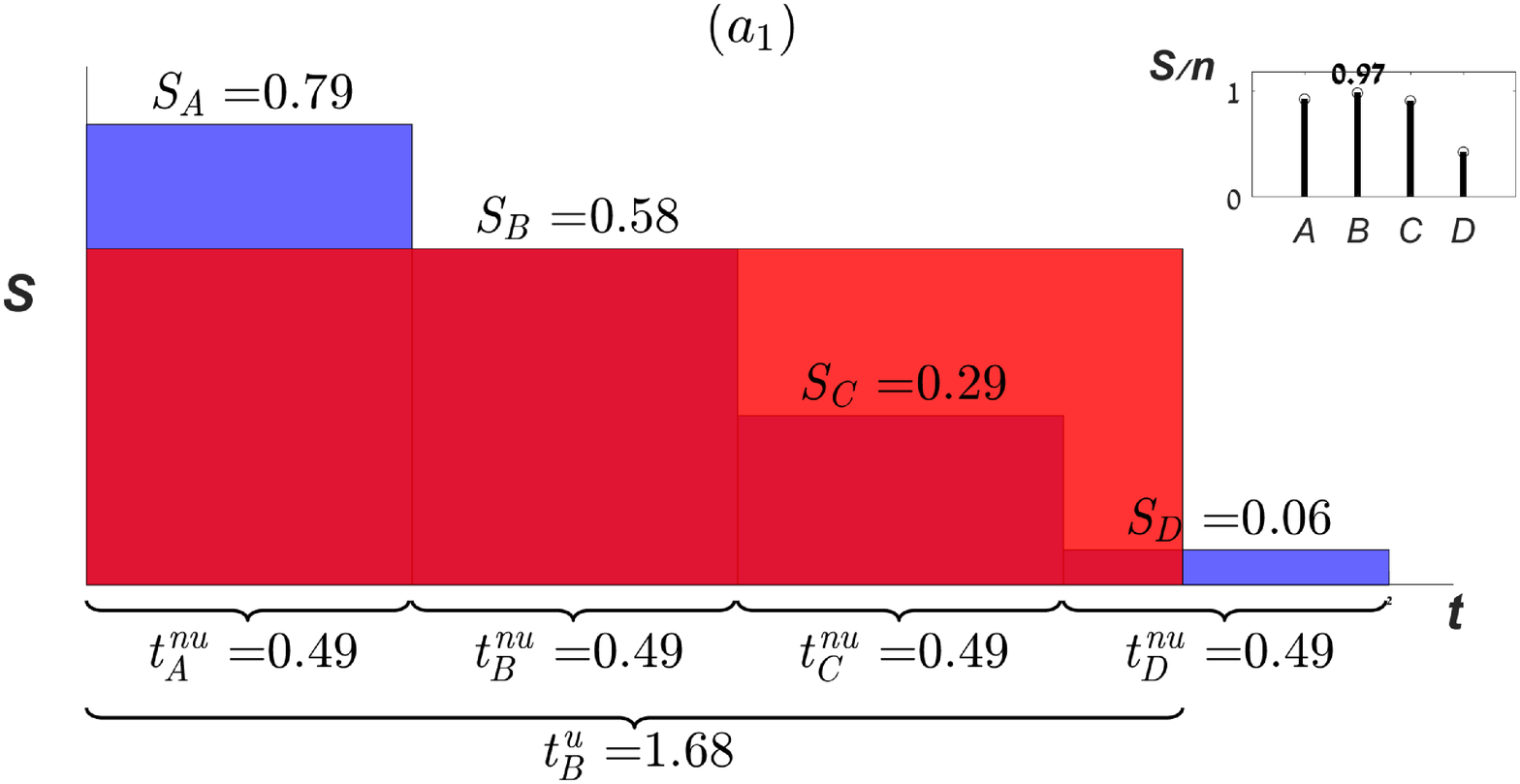}\hspace{-1em} &
	\includegraphics[width=10cm]{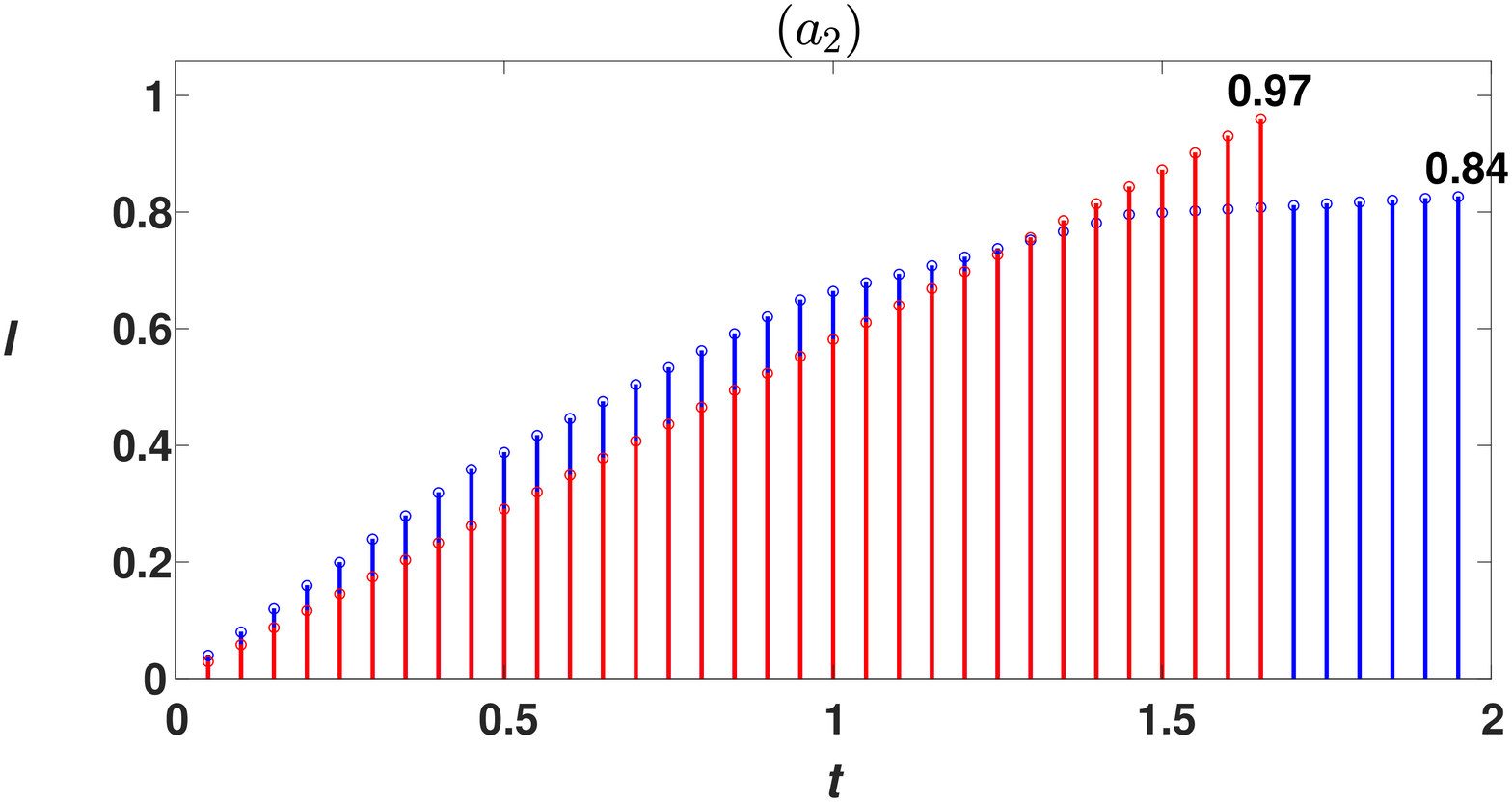} \\
	\hspace{-7em} \includegraphics[width=10cm]{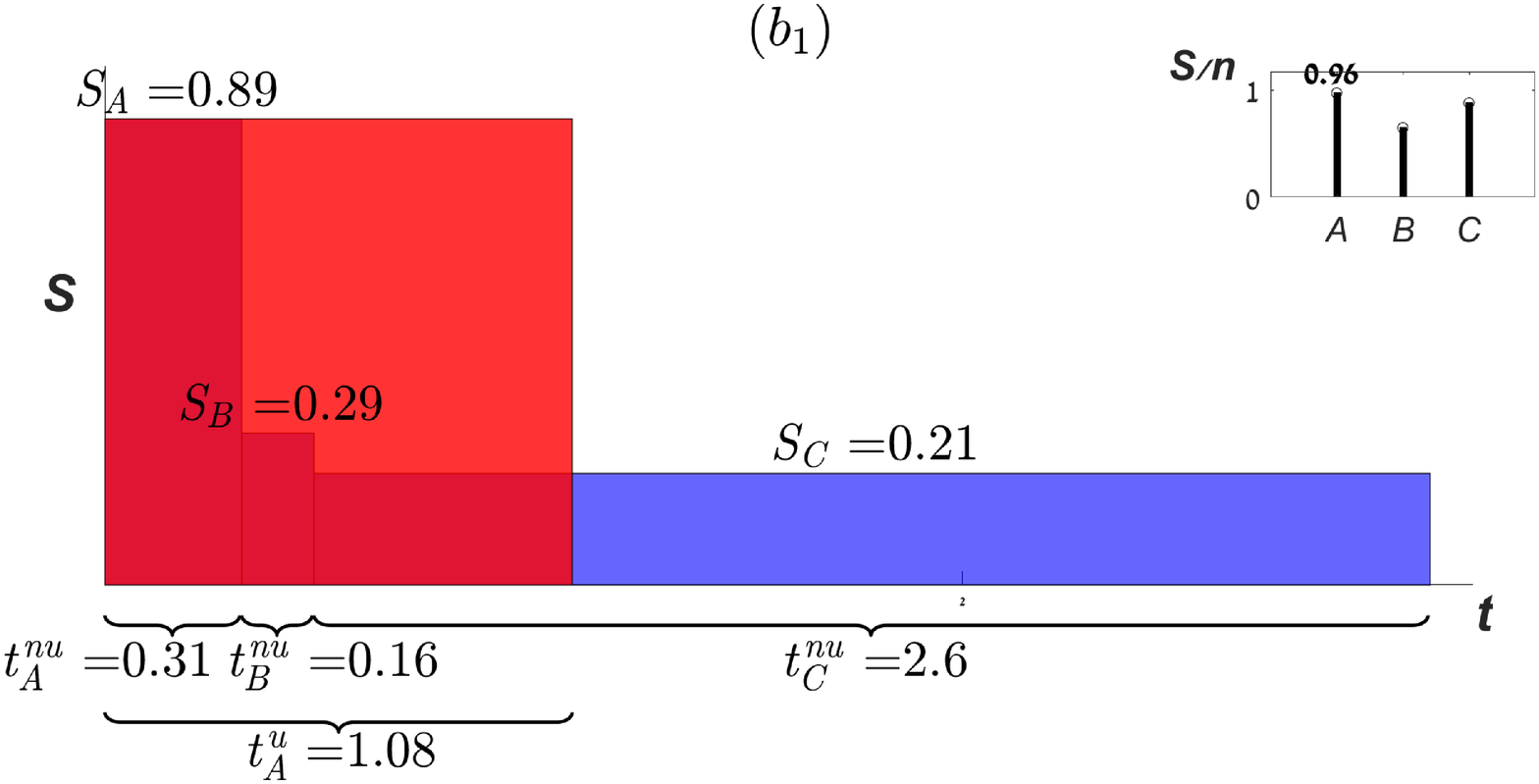}\hspace{-1em} &
	\includegraphics[width=10cm]{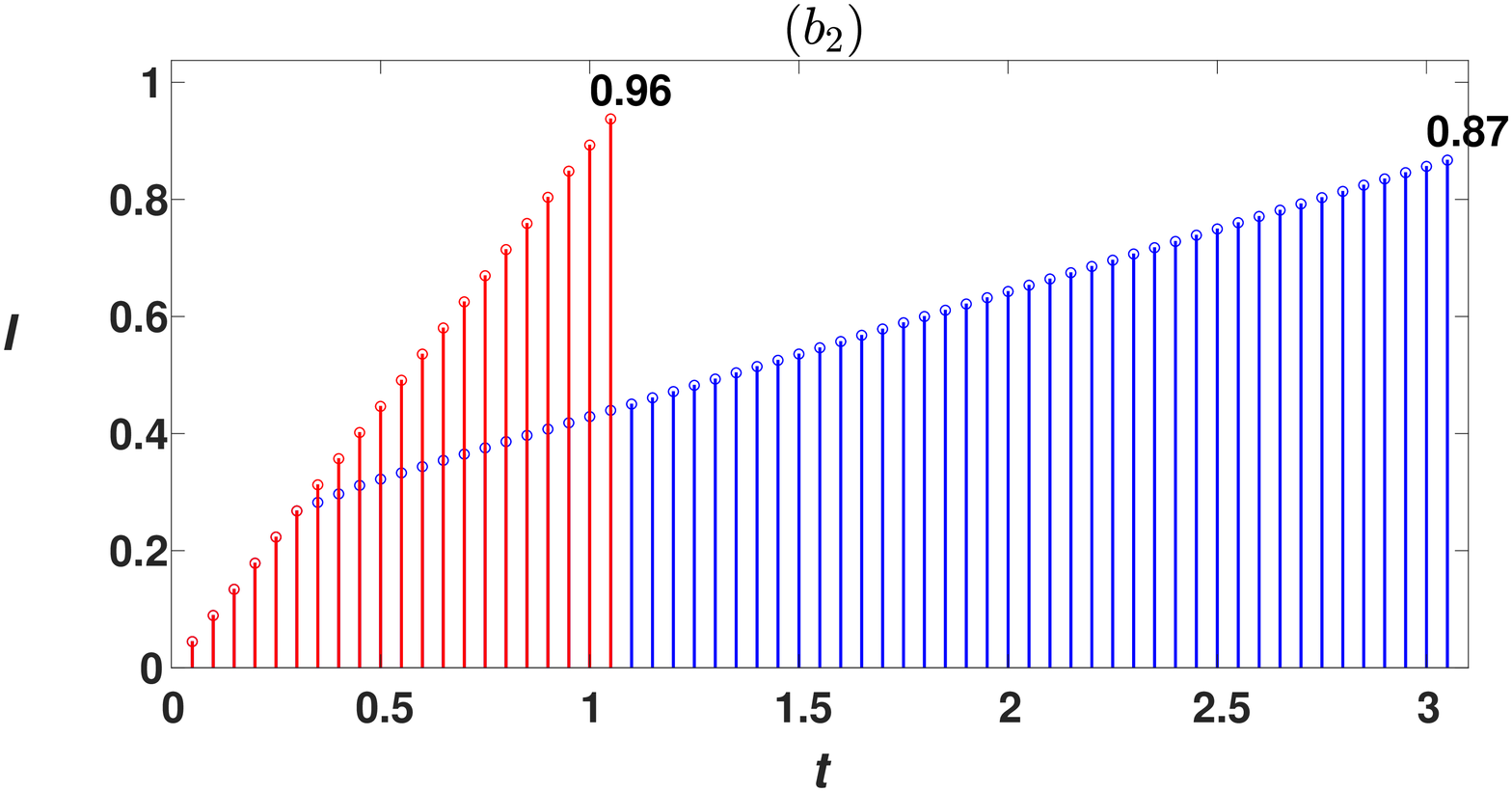} \\
\end{tabular}
\end{center}
\caption{\textbf{[($a$),($b$)] Failure stages of ER network: ($a$) Nonuniform lifetime division proportional to node degree:} Network size is $N=10^4$ nodes with average degree $\langle k\rangle=2$. \textbf{($a_1$)} Graph of giant component size vs. time. Four blue rectangles represent four failure stages with nonuniform lifetime division. Red rectangle represents single failure stage with uniform lifetime division. Giant component size is written on top of the relevant rectangle for each failure stage. Time duration is written and noted by braces, extended across the relevant rectangle, below the $t$ axis, for each failure stage. \textbf{$(a_2)$} Graph of accumulated network survivability vs. time. Blue bars and red bars represent nonuniform and uniform lifetime division, respectively. The diagram ends at a time point where the network collapses. For each division method, the total network survivability is written on top of its last graph point. \textbf{($b$) Nonuniform lifetime division according to percentages of total lifetime:} Network size is $N=10^4$ nodes with average degree $\langle k\rangle=2.5$. Percentages out of the total lifetime -- nodes with degree $k=1$ -- $10\%$, nodes with degrees $k=2,3$ -- $15\%$, and nodes with degrees equal or greater than $4$ -- $75\%$. \textbf{($b_1$)} Graph of giant component size vs. time, similar to ($a_1$). \textbf{($b_2$)} graph of accumulated network survivability vs. time, similar to ($b_1$). Averages were taken over $100$ realizations. For convenience, the averages were taken on networks were generated with maximum degree $k_{max}=10$. The total lifetime budget $T_0$ was taken to be $10^4$ similar to the network size, in order to normalize the network survivability to be between $0$ and $1$. Insets in \textbf{($a_1$)} and \textbf{($b_1$)}: bar graph of the ratio $\frac{S}{n}$ between the giant component size and the number of nodes, for each of the sets. Maximum for this ratio is written on top of the bar of the set on which this ratio is maximal.}
\label{fig:sim_ER}
\end{figure*}

\begin{figure*}[htbp]
\begin{center}
  \begin{tabular}{cc}
	\hspace{-7em} \includegraphics[width=10cm]{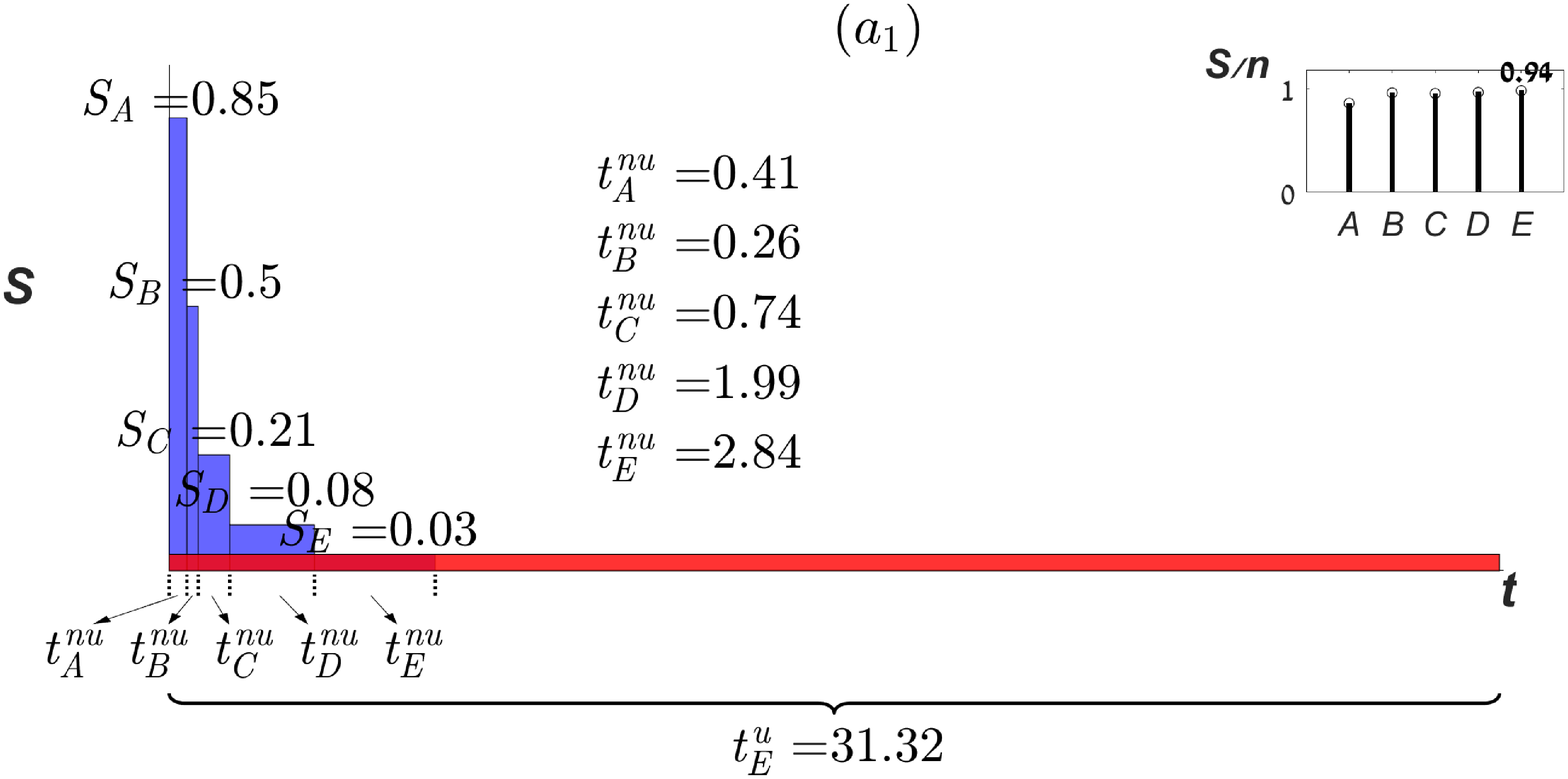}\hspace{-1em} &
	\includegraphics[width=10cm]{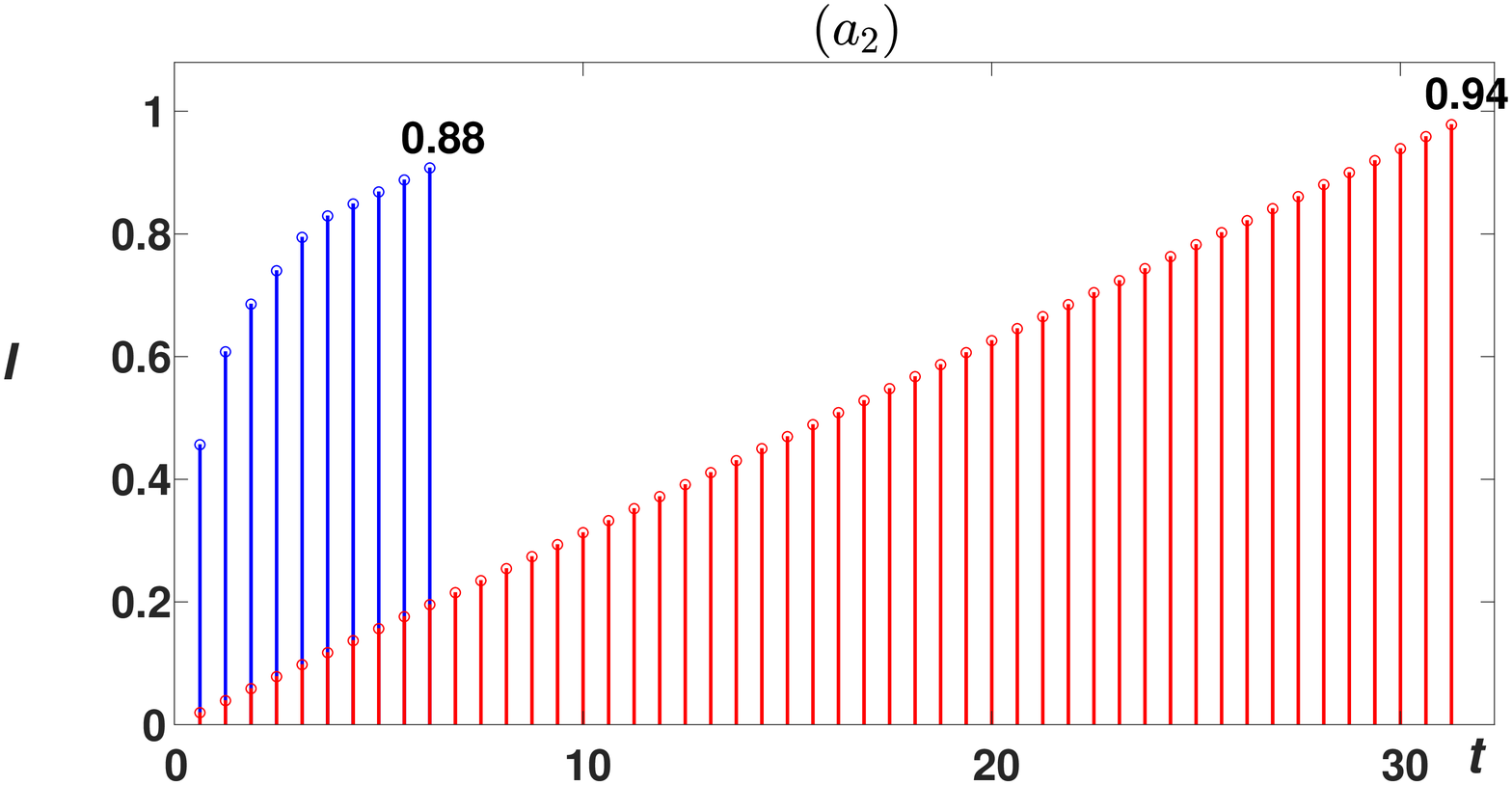} \\
	\hspace{-7em} \includegraphics[width=10cm]{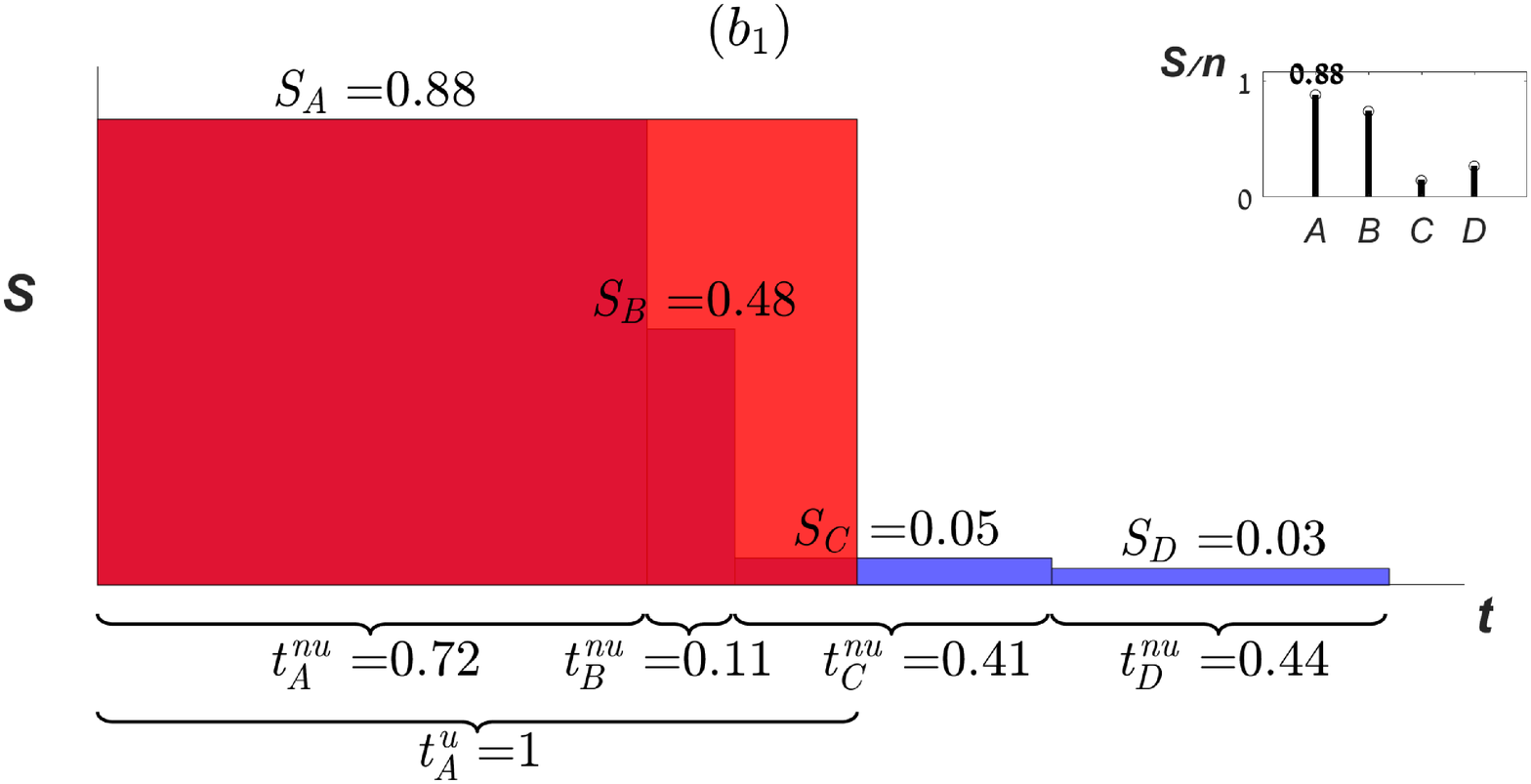}\hspace{-1em} &
	\includegraphics[width=10cm]{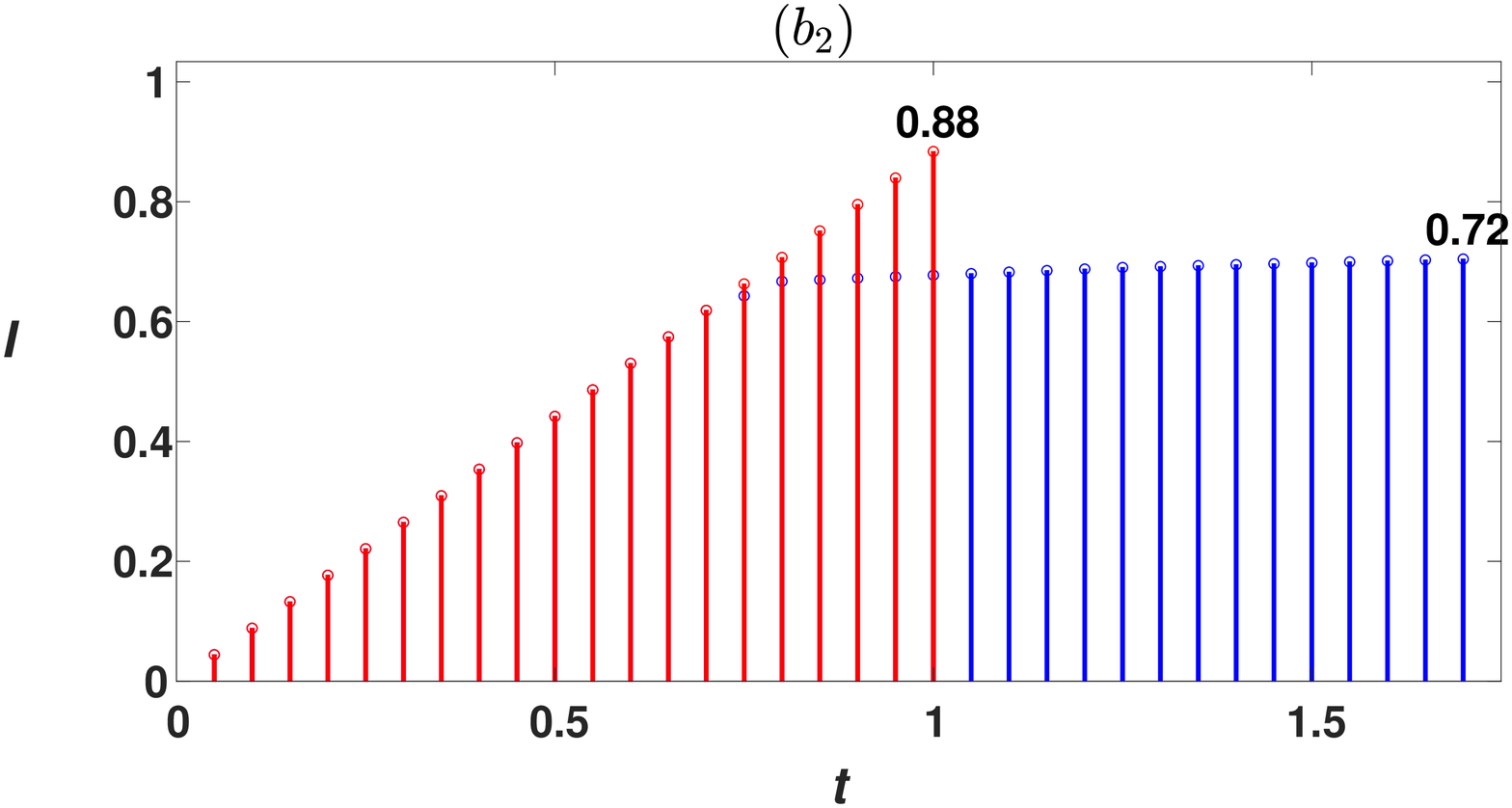} \\
\end{tabular}
\end{center}
\caption{\textbf{($a$) Failure stages of SF network:} Network size is $N=2.5\cdot10^4$ nodes with $\gamma=2.6$. Network was divided to five sets -- nodes with degree $1$, nodes with degree $2$, nodes with degrees $3$ and $4$, nodes with degrees between $5$ and $8$ and nodes with degrees equal or greater than $9$. Each set was allocated by $20\%$ of the total lifetime budget. \textbf{($a_1$)} Graph of giant component size vs. time. \textbf{($a_2$)} Graph of accumulated network survivability vs. time. Averages were taken over $100$ realizations. \textbf{($b$) Failure stages of Power-Grid real network:}  Network size is $N=4,941$ nodes. Only the first $85\%$ of nodes in the data set were considered. Percentages out of the total lifetime -- nodes with degree $k=1$ -- $30\%$, nodes with degrees $k=2$ -- $25\%$, nodes with degrees $k=3,4$ -- $25\%$, and nodes with degrees equal or greater than $5$ -- $20\%$. \textbf{($b_1$)} Graph of giant component size vs. time. \textbf{($b_2$)} Graph of accumulated network survivability vs. time. Insets in \textbf{($a_1$)} and \textbf{($b_1$)}: The ratio $\frac{S}{n}$ for each of the sets, similar to the insets of Figs. \ref{fig:sim_ER}($a_1$) and \ref{fig:sim_ER}($b_1$).}
\label{fig:sim_SF_Real}
\end{figure*}

Simulations were implemented on Erd\H{o}s-R\'{e}nyi (ER) and Scale-Free (SF) random networks, and on real networks. For each realization of each simulation of each of the above networks, firstly a budget of lifetime is distributed non-uniformly between the various nodes, due to a distribution rule. Due to that, the sets $A$, $B$, $C$ etc., as they were defined above in the model section, are identified. Then the network survivability simulation begins and a time clock is activated, whereby the different sets terminate their life according to their lifetimes $t_A$, $t_B$, $t_C$ etc. The time duration between two successive sets failures is defined as one stage of the whole network failure, and we note the last stage of the network failure by $m$. Then, for each one of the network failure stages, represented by $i$ ($i=1,2,3,...,m$), the product between the giant component size $S_{\Psi_i}$ and the current stage time duration $t_{\Psi_i}$ is calculated. The network survivability is calculated to be $I=\sum_{i=1}^mS_{\Psi_i}t_{\Psi_i}$.

In the second part of each simulation realization, our algorithm is activated. For each set of the network's nodes the ratio $\frac{S}{n}$ is calculated. We choose the set on which this ratio value is maximized, and divide the whole lifetime budget uniformly between this set's nodes. Then the network survivability simulation begins again and a time clock is activated. Obviously, the entire network collapses after the chosen set's lifetime is passed. The network survivability would be the product of the chosen set's giant component size by the chosen set's lifetime. 

Figures \ref{fig:sim_ER}($a_1$) and \ref{fig:sim_ER}($a_2$) present simulation results for ER network with $N=10^4$ nodes and mean degree $\langle k\rangle=2$. The nonuniform lifetime allocating rule is -- each node receives lifetime with proportion to its degree. A plot of the giant component size vs. the time during the network failure, is presented in Fig. \ref{fig:sim_ER}($a_1$), both for the nonuniform lifetime division (blue) and for our theory with uniform lifetime division (red). In the nonuniform lifetime division, the network survivability is spanned over four stages, each of them is presented by a blue rectangle. The giant component size for each stage, is written above the relevant rectangle, and the time duration of each stage is presented by braces below the $t$ axis, that are spanned over the stage's lifetime, and is written below the braces. For example, for the first stage $A$, $S_A=0.79$ and $t_A^{nu}=0.49$ (the superscript $^\prime nu^\prime$ notes nonuniform). Note that set $A$, in the first stage, includes all the network's nodes (except nodes with degree $0$ that obviously are not allocated with lifetime), set $B$, in the second stage after nodes with degree $k=1$ were failed (lifetime distribution with proportion to node's degree, causes the nodes with degree $k=1$ to receive minimum lifetime), includes nodes with degrees equal or greater than $2$, set $C$, in the third stage, includes nodes with degrees equal or greater than $3$ and accordingly set $D$ includes nodes with degrees equal or greater than $4$. The next stages are not presented in the figure, since when stage $D$ is terminated and nodes with degree $4$ are failed, the network is fragmented and the giant component is not exists anymore in the network. Hence, the contribution of these stages to the network survivability is $0$. Finally, the network survivability for nonunifrom lifetime allocation is $I=S_At_A^{nu}+S_Bt_B^{nu}+S_Ct_C^{nu}+S_Dt_D^{nu}$.

The uniform lifetime division according to our theory is presented by the red rectangle. As was described above, our algorithm calculates for each of the sets $A,B,C$ and $D$, the ratio $\frac{S}{n}$ of the giant component size and the number of nodes in the set. Our algorithm finds that the maximum for this ratio in our case, is obtained for set $B$. This result is shown in the inset of the figure, which is a bar graph with points for the sets $A,B,C$ and $D$ on the x axis, and a bar for each of the points whose height represents the ratio $\frac{S}{n}$ for the relevant set. We see that the highest bar belongs to set $B$ with a value of $0.97$. Accordingly, the lifetime budget is fully divided uniformly between set $B$ nodes, and the whole network fails after one stage only -- failure of set $B$ nodes. The giant component size in this only stage is written above the red rectangle $S_B=0.58$, and the time duration of this stage is $t_B^u=1.68$ (the superscript $^\prime u^\prime$ notes uniform) and is presented by braces below the $t$ axis. Obviously, the network survivability in this way is $I=S_Bt_B^u$.

Figure \ref{fig:sim_ER}$(a_2)$ is a graph of the network survivability $I$ vs. the time during the network failure, both for the nonuniform lifetime division (blue) and the uniform lifetime division (red). For each time $t$ on the $x$ axis, a bar is drawn whose height represents the value of the expression $\int_0^t S(t^\prime)dt^\prime$, which is the accumulated network survivability from the beginning of the network failure until time $t$. We can see that although in the nonuniform division the network survivability spans a longer time period than for the uniform division, the total survivability of the uniform division ($0.97$) is greater than for the nonuniform division ($0.84$). This result illustrates the idea of our theory, that for each proposal of nonuniform lifetime division between the network's nodes, there is other proposal of uniform lifetime division, on which the network survivability is greater.

Figures \ref{fig:sim_ER}$(b_1)$ and \ref{fig:sim_ER}$(b_2)$ present simulation results for ER network with $N=10^4$ nodes and mean degree $\langle k\rangle=2.5$. In this simulation, the nonuniform lifetime division was implemented according to percentages of the total lifetime, such that $10\%$ of the total lifetime was divided uniformly between the nodes with degree $1$, $15\%$ of the total lifetime was divided uniformly between the nodes with degrees $2$ and $3$, and the remained $75\%$ of the lifetime was divided uniformly between the nodes with degrees equal or greater than $4$. An interesting point is that according to this division, although the total lifetime budget for nodes with degree $1$ ($10\%$) is less than for nodes with degrees $2$ and $3$ ($15\%$), the lifetime of a single node with degree $1$ is greater than the lifetime of a single node with degree $2$ or $3$. This is because the network contains significantly more nodes with degrees $2$ and $3$ than nodes with degree $1$. As a result, set $A$ contains all the network's nodes except nodes with degree $0$, set $B$ contains nodes with degree $1$ and with degrees equal or greater than $4$, after the nodes with degrees $2$ and $3$ of set $A-B$ were failed, and set $C$ contains nodes with degrees equal or greater than $4$. 

Figure \ref{fig:sim_ER}$(b_2)$ validates again our theory. We can see that against the nonuniform lifetime division proposal with network survivability value of $0.87$, we propose a uniform lifetime division in which the network survivability is greater with value of $0.96$. 

Figures \ref{fig:sim_SF_Real}($a_1$) and \ref{fig:sim_SF_Real}($a_2$) present simulation results for SF network with $N=2.5\cdot10^4$ nodes and $\gamma=2.6$. The nonuniform lifetime division was implemented according to percentages of the total lifetime. We divide the network's nodes to five sets -- nodes with degree $1$, nodes with degree $2$, nodes with degrees $3$ and $4$, nodes with degrees between $5$ and $8$ and nodes with degrees equal or greater than $9$. Each set was allocated by $20\%$ of the total lifetime budget.
In Fig. \ref{fig:sim_SF_Real}($a_1$) we can see the interesting result, where our algorithm finds that maximum survivability would be achieved by uniformly lifetime division on set $E$ with the high degrees, that according to SF network properties contains a very small part of the network's nodes. This result can be explained due to the fact that the nodes in set $E$ have relatively high degrees, and the probability that all of them are connected together in one component is very high. Therefore, all, or at least most, of set $E$ nodes are part of the giant component of this set, and the ratio $\frac{S_E}{n_E}$ is very high, and would be greater than this ratio value for the other sets. Another interesting point is the very long time duration of network survivability with uniform lifetime division on set $E$, that is represented by the very long length of the red rectangle in Fig. \ref{fig:sim_SF_Real}($a_1$). This is due to the fact that on one hand set $E$ receives $20\%$ of the total lifetime, equally to the other sets, while on the other hand it contains a very small combination of nodes relative to the other sets. Therefore, each node in this set receives a relatively high lifetime amount, and this set's lifetime is very high.     

Figures \ref{fig:sim_SF_Real}($b_1$) and \ref{fig:sim_SF_Real}($b_2$) present simulation results for the Power-Grid (PG) network \cite{opsahl-url-networkdatasets}, as a demonstration of our theory on real networks too. Due to the nature of the nodes belonging to this network, which are electric elements, this network demonstrates a classical example for nodes removal due to aging. Since PG network is fully connected, then all its nodes are contained in its giant component, its ratio value $\frac{S}{n}$ is 1, and the solution for maximum survivability is very trivial -- dividing the total lifetime budget uniformly between all the network's nodes. In order to show nontrivial solutions for this network too, we choose to consider only part of the network's nodes, such that ignoring the other nodes causes the network to be not fully connected. Therefore, the results of Figs. \ref{fig:sim_SF_Real}($b_1$) and \ref{fig:sim_SF_Real}($b_2$) are related to PG network where only the first $85\%$ of the nodes in the data set were considered. The nonuniform lifetime division was implemented according to percentages of the total lifetime as follows: nodes with degree $1$ -- $30\%$, nodes with degree $2$ -- $25\%$, nodes with degrees $3$ and $4$ -- $25\%$, and nodes with degrees equal or greater than $5$ - $20\%$. The results in Fig. \ref{fig:sim_SF_Real}($b_2$) show again that also in PG real network, dividing the lifetime uniformly on set $A$ nodes -- the set that contains all the network's nodes except nodes with degree $0$, gives a network survivability value of $0.88$, that is greater than the survivability value of $0.72$ with the nonuniform lifetime division. 

\section{Summary}
In this work we developed a method for maximizing the robustness of a network in the dynamic approach, where the robustness measurement is performed by considering the network functionality during the entire nodes removal event, and according to a survivability function Eq. (\ref{eq:survivability}). We
proved analytically that for a given group of sets of nodes, all of them are possibly for lifetime allocation, in order to maximize the network survivability, we only have to allocate lifetime to the set for which the ratio between its giant component size and its number of nodes is maximal, where the allocation should be performed uniformly between the nodes of the chosen set. We hope that these findings could be useful at the stage of network design, as a tool for improving network survivability.\\

\bibliography{my_bib}

\clearpage

\end{document}